\documentclass[twocolumn,showpacs,aps]{revtex4}

\begin{document}

 \newcommand{\bq}{\begin{equation}}
 \newcommand{\eq}{\end{equation}}
 \newcommand{\bqn}{\begin{eqnarray}}
 \newcommand{\eqn}{\end{eqnarray}}
 \newcommand{\nb}{\nonumber}
 \newcommand{\lb}{\label}
\newcommand{\PRL}{Phys. Rev. Lett.}
\newcommand{\PL}{Phys. Lett.}
\newcommand{\PR}{Phys. Rev.}
\newcommand{\CQG}{Class. Quantum Grav.}

\title{Stability of the de Sitter spacetime in  Horava-Lifshitz theory}

\author{Yongqing Huang $^{a}$}
\email{yongqing_huang@baylor.edu}

\author{Anzhong Wang $^{a}$}
\email{anzhong_wang@baylor.edu}

\author{Qiang Wu $^{a, b}$}
\email{qiang_wu@baylor.edu}

\affiliation{$^{a}$ GCAP-CASPER, Physics Department, Baylor
University, Waco, TX 76798-7316, USA \\
$^{b}$Department of Physics, Zhejiang University of
Technology, Hangzhou 310032,  China }

\date{\today}

\begin{abstract}

 The stability of de Sitter spacetime in  Horava-Lifshitz theory of gravity with projectability
 but without detailed balance condition is studied. It is found that, in contrast to the case of
 the Minkowski background, the spin-0 graviton now is stable for  any given $\xi$, and free of
 ghost for $\xi \le 0$ in the infrared limit, where $\xi$ is the dynamical coupling constant.

\end{abstract}


\maketitle

\section{Introduction}
\renewcommand{\theequation}{1.\arabic{equation}} \setcounter{equation}{0}

Formulating a proper theory of quantum gravity has been one of the most challenging questions in
physics over the past several decades \cite{QG1,QG2}. Despite of innumerous efforts, so far there are only
two major candidates, the loop quantum gravity \cite{LQG} and strng/M theory \cite{string1,string2}. While the latter
seems to be  the best bet for such a theory that unifies all the known forces, it is often too complicated
to deal with.  The former, on the other hand, has its own challenging problems.

Recently, Horava proposed a theory of quantum gravity \cite{Horava1,Horava2,Horava3}, motivated by the Lifshitz theory in
solid state physics \cite{Lifshitz}. The Horava-Lifshitz (HL) theory 
 is based on the perspective
that Lorentz symmetry should appear as an emergent symmetry at long distances, but can be fundamentally
absent at high energies \cite{Pav1,Pav2}.   At low energies, the theory is expected to  flow to GR,   whereby the Lorentz
invariance is ``accidentally restored." The theory is   non-relativistic, ultra-violet (UV) complete, explicitly breaks
Lorentz invariance at short distances, but is expected to recover GR in the  infrared (IR) limit.

The effective speed of light in this  theory diverges in the UV regime, which could potentially resolve the
horizon problem without invoking inflation \cite{KK}. The  spatial curvature is enhanced by higher-order
curvature terms \cite{calcagni1,calcagni2,LMP,WM}, and this opens a new approach to the flatness problem and to
a bouncing universe \cite{calcagni1,calcagni2,brand,WW}. In addition,  in the super-horizon region scale-invariant
curvature perturbations can be produced without inflation \cite{Muk1,Muk2,Muk3,WMW}, and  the perturbations
become adiabatic during slow-roll inflation driven by a single scalar field and the comoving curvature perturbation
is constant  \cite{WMW}. Due to all  these remarkable features, the theory has attracted lot of  attention
lately \cite{BHs1,BHs2,BHs3,BHs4,BHs5,BHs6,BHs7,BHs8,BHs9,BHs10,BHs11,BHs12,BHs13,BHs14,BHs15,%
BHs16,BHs17,BHs18,BHs19,BHs20,BHs21,BHs22,BHs23,BHs24,BHs25,BHs26,BHs27,BHs28,BHs29,BHs30,BHs31,%
BHs32,BHs33,BHs34,BHs35,BHs36,BHs37,BHs38,BHs39,BHs40,BHs41,BHs42,BHs43,BHs44,BHs45,BHs46,%
Cosmos1,Cosmos2,Cosmos3,Cosmos4,Cosmos5,Cosmos6,Cosmos7,Cosmos7.1,Cosmos8,Cosmos8.1,Cosmos9,Cosmos10,%
Cosmos11,Cosmos12,Cosmos13,Cosmos14,Cosmos15,Cosmos16,Cosmos17,Cosmos18,Cosmos19,Cosmos20,%
Cosmos21,Cosmos22,Cosmos23,Cosmos24,Cosmos25,Cosmos26,Cosmos27,Cosmos28,Cosmos29,Cosmos30,Cosmos31,%
others1,others1.1,others2,others3,others4,others5,others6,others7,others8,others9,others10,others11,others12,%
others13,others14,others15,others16,others17,others18,others19,others20,others21,others22,others23,others24,%
others25,others26,others27,others28,others29,others30,others31,others32,others33,others34,others35,others36,%
others37,others38,others39,others40,others41,others42,others43,others44,others45,others46,others47,others48,others49}.

In the HL theory,  it was assumed  two conditions -- {\em detailed balance and projectability}   \cite{Horava1,Horava2,Horava3}.
The detailed balance condition restricts the form of a general potential in a ($D+1$)-dimensional Lorentz action
to a specific form that can be expressed in terms of a D-dimensional action of a relativistic theory with Euclidean
signature, whereby the number of independent-couplings is considerably limited. The projectability condition, on
the other hand, originates from the fundamental symmetry of the theory -- the foliation-preserving
diffeomorphisms of the Arnowitt-Deser-Misner  (ADM) form, and is crucial to form a closed set of Poisson
brackets \cite{LP1,LP2}.

 So far most of the work  has  abandoned the projectability condition but  kept the detailed
balance \cite{BHs1,BHs2,BHs3,BHs4,BHs5,BHs6,BHs7,BHs8,BHs9,BHs10,BHs11,BHs12,BHs13,BHs14,BHs15,%
BHs16,BHs17,BHs18,BHs19,BHs20,BHs21,BHs22,BHs23,BHs24,BHs25,BHs26,BHs27,BHs28,BHs29,BHs30,BHs31,%
BHs32,BHs33,BHs34,BHs35,BHs36,BHs37,BHs38,BHs39,BHs40,BHs41,BHs42,BHs43,BHs44,BHs45,BHs46,%
Cosmos1,Cosmos2,Cosmos3,Cosmos4,Cosmos5,Cosmos6,Cosmos7,Cosmos7.1,Cosmos8,Cosmos8.1,Cosmos9,Cosmos10,%
Cosmos11,Cosmos12,Cosmos13,Cosmos14,Cosmos15,Cosmos16,Cosmos17,Cosmos18,Cosmos19,Cosmos20,%
Cosmos21,Cosmos22,Cosmos23,Cosmos24,Cosmos25,Cosmos26,Cosmos27,Cosmos28,Cosmos29,Cosmos30,Cosmos31,%
others1,others1.1,others2,others3,others4,others5,others6,others7,others8,others9,others10,others11,others12,%
others13,others14,others15,others16,others17,others18,others19,others20,others21,others22,others23,others24,%
others25,others26,others27,others28,others29,others30,others31,others32,others33,others34,others35,others36,%
others37,others38,others39,others40,others41,others42,others43,others44,others45,others46,others47,others48,others49}. 
However, with detailed
balance a scalar field is not UV stable \cite{calcagni1,calcagni2}, and gravitational perturbations in the scalar section
have ghosts  \cite{Horava1,Horava2,Horava3} and are not stable for any given value of the dynamical coupling constant
$\xi \; (\equiv 1 - \lambda)$ \cite{BS}.   In addition, detailed balance also requires a non-zero (negative)
cosmological constant, breaks the parity in the purely gravitational sector \cite{SVW1,SVW2,SVW3}, and makes the
perturbations not scale-invariant \cite{GWBR}. Breaking the  projectability condition, on the other hand,
can cause strong couplings  \cite{CNPS1,CNPS2,CNPS3,CNPS4} and gives rise to an inconsistency theory \cite{LP1,LP2}.

To resolve these problems, various modifications have been proposed \cite{NewHLs1,NewHLs2,NewHLs3,NewHLs4,%
NewHLs5,NewHLs6,NewHLs7,NewHLs8,NewHLs9}. In particular,
Blas, Pujolas and Sibiryakov (BPS) \cite{BPS1,BPS2} showed that the strong coupling problem can be solved
without projectability condition 
when terms constructed from the  gradient of the lapse function 
are included. However, as shown in \cite{PS}, strong coupling may still exist, unless the scale appearing in front of
the higher order terms is much lower than the Planck scale \cite{BPS1,BPS2}. It is also not clear how the inconsistency
problem found in \cite{LP1,LP2} is resolved in such a setup.  Moreover, in the IR limit the theory is identical to a special
case of the Einstein-aether theory, where the vector field is hypersurface-orthogonal \cite{Jaconson}. But, the latter
seemingly conflicts with observations \cite{Zuntz}.

On the other hand, Sotiriou, Visser and Weinfurtner (SVW) formulated the most general HL theory with
projectability condition but without the detailed balance \cite{SVW1,SVW2,SVW3}. Writing the 4-dimensional metric
in  the  ADM form,
 \bqn
 \lb{1.2}
ds^{2} &=& - N^{2}c^{2}dt^{2} + g_{ij}\left(dx^{i} + N^{i}dt\right)
     \left(dx^{j} + N^{j}dt\right), \nb\\
     & & ~~~~~~~~~~~~~~~~~~~~~~~~~~~~~~  (i, \; j = 1, 2, 3),~~~
 \eqn
the projectability condition requires that
\bq
\lb{1.4}
N = N(t), \;\;\; N^{i} = N^{i}(t, x),\;\;\; g_{ij} = g_{ij}(t, x).
\eq
Note that in \cite{WM,WMW,GPW}, the constant $c$ representing  the speed of light  was absorbed into $N$.
The ADM form (\ref{1.2}) is preserved only by  the types of coordinate transformations,
\bq
\lb{1.3}
t \rightarrow f(t),\; \;\; x^{i} \rightarrow \zeta^{i}(t, {\bf x}).
\eq
Due to these restricted diffeomorphisms, one more degree of freedom appears
in the gravitational sector - a spin-0 graviton. This is potentially dangerous, and needs to be highly
suppressed in the IR regime, in order to be consistent with observations. Similar problems also raise
in other modified theories, such as massive gravity \cite{RT1}.

Then, it can be shown that the most general action, which  preserves the parity and is with projectability
but without detailed balance condition,  is given by \cite{SVW1,SVW2,SVW3},
 \bqn \lb{1.6}
S = \zeta^2\int dt d^{3}x N \sqrt{g} \left({\cal{L}}_{K} -
{\cal{L}}_{{V}}+\zeta^{-2} {\cal{L}}_{M} \right),
 \eqn
where $g={\rm det}\,g_{ij}$, $ {\cal{L}}_{M}$ denotes the matter Lagrangian density,
and
 \bqn \lb{1.7}
{\cal{L}}_{K} &=& K_{ij}K^{ij} - \left(1-\xi\right)  K^{2},\nb\\
{\cal{L}}_{{V}} &=& 2\Lambda - R + \frac{1}{\zeta^{2}}
\left(g_{2}R^{2} +  g_{3}  R_{ij}R^{ij}\right)\nb\\
& & + \frac{1}{\zeta^{4}} \left(g_{4}R^{3} +  g_{5}  R\;
R_{ij}R^{ij}
+   g_{6}  R^{i}_{j} R^{j}_{k} R^{k}_{i} \right)\nb\\
& & + \frac{1}{\zeta^{4}} \left[g_{7}R\nabla^{2}R +  g_{8}
\left(\nabla_{i}R_{jk}\right)
\left(\nabla^{i}R^{jk}\right)\right],
 \eqn
where $\zeta^{2} = 1/{16\pi G}$, and 
the covariant derivatives and
Ricci and Riemann terms are all constructed from the three-metric $g_{ij}$,
while $K_{ij}$ is the extrinsic curvature,
 \bq \lb{1.8}
K_{ij} = \frac{1}{2N}\left(- \dot{g}_{ij} + \nabla_{i}N_{j} +
\nabla_{j}N_{i}\right),
 \eq
where $N_{i} = g_{ij}N^{j}$. The constants $\xi, g_{I}\,
(I=2,\dots 8)$  are coupling constants, and $\Lambda$ is the
cosmological constant. In the IR limit, all the high order curvature terms (with
coefficients $g_I$) drop out, and the total action reduces when
$\xi = 0$ to the Einstein-Hilbert action.

It should be noted that, although the SVW generalization  seems very promising,  the gravitational scalar perturbations
in such a setup either have  ghosts ($0 \le \xi \le 2/3$) or are not  stable ($\xi < 0$) \cite{WM,KA}. 
In order to avoid ghost instability, one needs to assume $\xi \le 0$. Then, the sound speed
$c_{\psi}^{2} = \xi/(2-3\xi)$ becomes imaginary, which leads to an IR instability. Izumi and Mukohyama
showed that this type of instability does not show up if $|c_{\psi}|$ is less than a critical value \cite{IM}.

In this brief report, we show explicitly that this is no longer the case in the de Sitter background. The gravitational scalar perturbations
are stable for any given $\xi$, and are free of  ghosts  for $\xi \le  0$. In particular, in the next section we shall give a brief review
of scalar perturbations in a flat FRW background, while in Section III we consider perturbations in a de Sitter background. The
paper is ended with Section IV, where our main conclusions are presented.

It should be noted that gravitational scalar perturbations in the HL theory with detailed balance was studied in \cite{CHZ}, while
the stability of the Einstein static universe was considered in \cite{BL1,BL2}.

\section{Scalar Perturbations in  Flat FRW Backgrounds}

\renewcommand{\theequation}{2.\arabic{equation}} \setcounter{equation}{0}

We give a very brief introduction to the scalar perturbations of flat FRW background in the HL gravity without detailed
balance, but with the projectability condition. For  detail, we refer readers to \cite{WM,WMW}.
The homogeneous and isotropic flat universe is described by the FRW
metric, $ ds^{2} = a^{2}(\eta)\left(- d\eta^{2} + \delta_{ij}dx^{i}dx^{j}\right)$.
For this metric, $\bar K_{ij} = - a {\cal{H}} \delta_{ij}$ and $\bar R_{ij} =0$, where
${\cal{H}} = {a'}/a$ and an  overbar denotes a background quantity. Then,
  the Hamiltonian constraint yields,
 \bq \lb{3.4a}
\left(1 - \frac{3}{2}\xi\right)\frac{{\cal{H}}^{2}}{a^{2}} =
\frac{8\pi G}{3}\bar \rho+ \frac{\Lambda}{3},
 \eq
while the dynamical equations  give rise to
 \bq \lb{3.4b}
\left(1 - \frac{3}{2}\xi\right)\frac{{\cal{H}}'}{a^{2}} =  - {4\pi G\over
3}(\bar\rho+3\bar p)+ {1\over3} \Lambda,
 \eq
where $\bar\rho$ and $\bar p$ denote the energy density and pressure of matter of the FRW background.
Similarly to GR, the super-momentum constraint
  is then satisfied identically, while the conservation laws of energy and momentum yield,
   \bq \lb{3.4e}
{\bar{\rho}}' + 3{\cal{H}} \left(\bar\rho +\bar p \right) = 0.
 \eq
This can be also obtained from Eqs.~(\ref{3.4a}) and (\ref{3.4b}). Clearly, replacing
$G$ and $\Lambda$, respectively, by $G/(1 - 3\xi/2)$ and $\Lambda/(1 - 3\xi/2)$,
Eqs.~(\ref{3.4a}) and (\ref{3.4b}) reduce exactly to the ones given in GR.

Linear scalar perturbations of the metric are given by
 \bqn \lb{4.1}
\delta{g}_{ij} &=&   a^{2}(\eta)\left(- 2\psi\delta_{ij} + 2E_{,ij}\right),\nb\\
\delta{N}_{i} &=&   a^{2}(\eta) B_{,i}   ~~ \delta{N} = a(\eta) \phi(\eta).
 \eqn
Choosing the quasi-longitudinal gauge \cite{WM},
\bq
\lb{4.2}
\phi = 0 = E,
\eq
we find that the two gauge-invariant quantities defined in \cite{WM} reduce to,
\bq
\lb{4.3}
\Phi = {\cal{H}}B + B',\;\;\;
\Psi = \psi - {\cal{H}}B,
\eq
and that to second order the actions take the forms,
\bqn
\lb{4.4a}
S^{(2)}_{K} &=& \zeta^{2}\int{d\eta d^{3}x a^{2}\Bigg\{\big(3\xi - 2\big)\Bigg[3\psi'^{2} + 6{\cal{H}}\psi\psi' + 2 \psi' \partial^{2}B}\nb\\
& & ~~~~~~~~~~~~~ ~~~~~~~~  + \frac{9}{2}{\cal{H}}^{2}\psi^{2}\Bigg] + \xi B\partial^{4}B\Bigg\},\\
\lb{4.4b}
S^{(2)}_{V} &=& \zeta^{2}\int{d\eta d^{3}x a^{2}\Bigg\{2\big(\partial \psi\big)^{2} - 3 \Lambda a^{2}\psi^{2}
- \frac{2\alpha_{1}}{a^{2}}\big(\partial^{2}\psi\big)^{2}}\nb\\
&& ~~~~~~~~~~~~~~~~~~~~~~ + \frac{2\alpha_{2}}{a^{4}}\psi \partial^{6}\psi\Bigg\},
\eqn
 where $ \alpha_{1} \equiv \big(8g_{2} + 3g_{3}\big)/\zeta^{2}$ and $ \alpha_{2} \equiv \big(3g_{8} - 8g_{7}\big)/\zeta^{4}$.

The matter perturbations are written as
 \bqn
 \lb{4.5}
\delta J^t &=& -2\delta \mu,\;\;\; \delta{J}^{i} = \frac{1}{a^{2}}q^{,i}\nb\\
\delta{\tau}^{ij} &=& \frac{1}{a^{2}}\Big[\left(\delta{\cal P} +
2\bar{p}\psi\right)\,\gamma^{ij} + {\Pi}^{,\langle ij\rangle}\Big].
 \eqn
The angled brackets on indices define the trace-free part:
 \bq
 \lb{4.6}
f_{,\langle ij \rangle} \equiv f_{,ij}-{1\over 3}
\delta_{ij}f_{,k}{}^{,k}.
 \eq
In GR, $\delta\mu$ reduces to the density perturbation $\delta\rho$, and
$q,\; \delta{\cal{P}},\; \Pi$ to, respectively, $ - a(\bar\rho+\bar p)\big(v + B\big)$, the
pressure perturbation $\delta{p}$, and the scalar mode of the anisotropic pressure.

 To first-order the Hamiltonian constraint is
 \bq \lb{4.7}
 \int d^{3}x\Bigg[\partial^{2}\psi
- \left(1 -\frac{3}{2}\xi\right){\cal H}
\left(\partial^2 B + 3\psi'\right)   -{4\pi G a^{2}}\delta{\mu}\Bigg]=0.
 \eq
The integrand is a generalization of the
Poisson equation in GR \cite{MW09}.
The supermomentum constraint, on the other hand, reads
 \bq \lb{4.8}
\left(2 - 3\xi\right){\psi}'  = {\xi}
 \partial^{2}B +  8\pi G a\, {q},
 \eq
which generalizes the general relativity $0i$ constraint
\cite{MW09}.
The trace and trace-free parts of the perturbed dynamical equations     yield, respectively,
  \bqn \lb{4.9a}
{\psi}'' &+& 2{\cal H}{\psi}' - {\xi \over 2-3\xi}\left(1+
\frac{\alpha_1}{a^2}\partial^2 +\frac{\alpha_2}{a^4}\partial^4
\right) \partial^2\psi
\nb\\
&=&  {8\pi Ga^{2} \over 3(2-3\xi)}\Bigg[3\delta{\cal P} + (2-3\xi)
\partial^{2}\Pi\Bigg],\\
  \label{4.9b}
 \left(a^{2}B\right)'  &=& \left(a^{2} + \alpha_{1} \partial ^{2}
+  \frac{\alpha_{2}}{a^{2}}\partial ^{4}\right) \psi  -8\pi
Ga^{4} \Pi.~~~~
 \eqn
The conservation laws    give
  \bqn \lb{4.10a}
& & \int d^{3}x \Big[\delta\mu' + 3{\cal H} \left(\delta{\cal P} +
\delta\mu\right)  -3 \left(\bar\rho +
\bar p\right){\psi}' \Big] = 0,\;\;\;~~~~\\
\lb{4.10b} & &   q'+3{\cal H}q     = a\delta{\cal P} +
{2\over3}a\partial^{2}\Pi. ~~~~
 \eqn

\section{Stability of the de Sitter Spacetime in the IR Limit}
\renewcommand{\theequation}{3.\arabic{equation}} \setcounter{equation}{0}

To see how the ghost and instability problems of scalar perturbations are avoided in the
de Sitter background, it is instructive first to recall how they raise in the Minkowski background
\cite{WM,KA}. Since in this section we are mainly concerned with IR limit, the terms proportional to
$\alpha_{1}$ and $\alpha_{2}$ are highly suppressed by the Planck scales $M^{2}_{pl}$ and
$M^{4}_{pl}$, respectively. Then, in the following discussions it is quite safe to neglect all these terms.

In the Minkowski background, without matter perturbations, Eq. (\ref{4.8}) in the momentum space
gives,
\bq
\lb{5.1}
k^{2}B_{k} = \frac{3\xi -2}{\xi} \psi_{k}',
\eq
for $\xi \not= 0$. Then, Eq.(\ref{4.9a}) becomes
\bq
\lb{5.2}
\frac{1}{c^{2}_{\psi}} \psi_{k}'' + k^{2}
\psi_{k} = 0,
\eq
where $c^{2}_{\psi} \equiv \xi/(2-3\xi)$. Clearly, it is stable 
only when $c^{2}_{\psi} \ge 0$,
that is, $ 0 < \xi \le 2/3$. However, submitting Eq. (\ref{5.1}) into Eqs. (\ref{4.4a}) and (\ref{4.4b}),
we find that
\bq
\lb{5.3}
{\cal{L}} \equiv {\cal{L}}_{K} - {\cal{L}}_{V}
= - \Bigg(\frac{1}{c^{2}_{\psi}} {\psi'}^{2} - \big(\partial\psi\big)^{2}\Bigg).
\eq
Therefore, unless $c^{2}_{\psi} < 0$ (or $\xi < 0$), the spin-0 graviton is a ghost.
But, when $\xi < 0$ the scalar field becomes unstable. Note that the spin-0
graviton becomes stable when $\xi = 0$ \cite{WM}.

The de Sitter spacetime is given by $a(\eta) = -1/(H\eta)$, where $\eta \le 0$. In particular,
 $\eta = - \infty$ corresponds to the initial ($t = 0$) of the universe, while $\eta = 0^{-}$ to the future infinity
 ($t = \infty$).  When matter is not present, we have
\bq
\lb{5.4}
q = \delta{\cal{P}} = \delta\mu = \Pi = 0,
\eq
 and the momentum constraint (\ref{4.8}) yields the same equation (\ref{5.1}) for $\xi \not= 0$. Then,
 from Eqs. (\ref{4.4a}) and (\ref{4.4b}), we find that
\bqn
\lb{5.5}
{\cal{L}}
&=& \alpha^{2}\Bigg\{\frac{2(3\xi - 2)}{\xi} \hat{\psi}_{k}'^{2} + \Bigg[\frac{9}{2}(2-\xi)(3\xi - 2){\cal{H}}^{2} \nb\\
& & ~~~~~~~
- 2k^{2}\Bigg]
{\hat{\psi}_{k}}^{2}\Bigg\},
\eqn
where
\bq
\lb{5.6}
{\psi}_{k} = \alpha \hat{\psi}_{k},\;\;\; \alpha = \frac{\alpha_{0}}{a^{3\xi/2}},
\eq
with $\alpha_{0}$ being an arbitrary constant. Thus, to have the kinetic part non-negative, we must assume
either $\xi \le 0$ or $\xi \ge 2/3$. However, the GR limit requires $\xi = 0$. Therefore, one needs to restrict to
the range $\xi \le 0$.  But, in the following we shall leave this possibility open, and show that the de Sitter
spacetime is stable against gravitational scalar perturbations for any given $\xi$ in the IR limit. 
To this purpose, we first notice that Eq. (\ref{4.9a})  can be cast in the form,
\bq
\lb{5.7}
\chi_{k}'' + \Bigg(\frac{\xi k^{2}}{2-3\xi} - \frac{2}{\eta^{2}}\Bigg)\chi_{k} = 0,
\eq
where $\chi_{k} = a\psi_{k}$. Depending on the values of $\xi$, the above equation has different solutions.
In the following we consider them separately.

{\bf Case 1) $\; \xi/(2-3\xi) < 0$:} In this case, Eq. (\ref{5.7}) has the general solution,
\bq
\lb{5.8}
\chi_{k} = c_{1}\Big(1 - \frac{1}{z}\Big)e^{z} + c_{2}\Big(1 + \frac{1}{z}\Big)e^{-z},
\eq
where $c_{1}$ and $c_{2}$ are two integration constant, and
\bq
\lb{5.9}
z \equiv \left|\frac{\xi k^{2}}{2 - 3\xi}\right|^{1/2} \eta = z_{0}\eta.
\eq
Then, $\psi_{k}$ and $B_{k}$ are given by
\bqn
\lb{5.10}
\psi_{k} &=& \tilde{c}_{1}\big(z - 1\big)e^{z} +  \tilde{c}_{2}\big(z+1\big)e^{-z},\nb\\
B_{k} &=& \frac{(3\xi -2)z}{\xi k^{2}} \Big(\tilde{c}_{1} e^{z} -  \tilde{c}_{2} e^{-z}\Big),
\eqn
which are all finite as $k\eta \rightarrow 0^{-}$ or $t \rightarrow \infty$, where
$\tilde{c}_{i}  \equiv - Hc_{i}/z_{0}$.  Inserting the above into
Eq.(\ref{4.3}), we find that the gauge-invaraint quantities $\Phi$ and $\Psi$ are given by
\bqn
\lb{5.11}
\Phi_{k} &=&  \frac{z_{0}z(3\xi -2)}{\xi k^{2}} \Big(\tilde{c}_{1} e^{z} +  \tilde{c}_{2} e^{-z}\Big),\nb\\
\Psi_{k} &=& \tilde{c}_{1}\Bigg[z + \frac{z_{0}(3\xi - 2) - \xi k^{2}}{\xi k^{2}} \Bigg] e^{z}\nb\\
& & + \tilde{c}_{2}\Bigg[z - \frac{z_{0}(3\xi - 2) - \xi k^{2}}{\xi k^{2}} \Bigg] e^{-z},
\eqn
which are also finite in the IR limit $k\eta \rightarrow 0^{-}$.

{\bf Case 2) $\; \xi/(2-3\xi) > 0$:} In this case, Eq. (\ref{5.7}) has the general solution,
\bq
\lb{5.12}
\chi_{k} = c_{1}\sin\big(z+c_{2}\big) + \frac{c_{1}}{z}\cos\big(z+c_{2}\big),
\eq
while  $\psi_{k},\; B_{k}$ and the gauge-invaraint quantities $\Phi$ and $\Psi$ are given, respectively, by
\bqn
\lb{5.13}
\psi_{k} &=& \tilde{c}_{1}\Big[z\sin\big(z+c_{2}\big) +  \cos\big(z+c_{2}\big)\Big], \nb\\
B_{k}&=& \frac{\tilde{c}_{1} z_{0}(3\xi -2)}{\xi k^{2}} z \cos\big(z+c_{2}\big),\nb\\
\Phi_{k} &=&  -\frac{\tilde{c}_{1} z_{0}^{2}(3\xi -2)}{\xi k^{2}}z \sin\big(z+c_{2}\big),\nb\\
\Psi_{k} &=& \frac{\tilde{c}_{1}}{\xi k^{2}}\Bigg[\xi k^{2} +  z_{0}^{2}(3\xi -2)\Bigg]\cos\big(z+c_{2}\big)\nb\\
& & ~~  + \tilde{c}_{1}z \sin\big(z+c_{2}\big),
\eqn
which in the IR limit $k\eta \rightarrow 0^{-}$ are finite, too.

{\bf Case 3) $\; \xi = 0$:} In this case, Eq. (\ref{4.8}) yields $\psi_{k} = \psi_{k}^{0}$, where $\psi_{k}^{0}$
is a constant, while Eq. (\ref{4.9b}) gives
\bq
\lb{5.14}
B_{k} =  \psi_{k}^{0}\eta + c_{0} H^{2}\eta^{2},
\eq
where $c_{0}$ is another integration constant. Clearly, both of these two terms represent decaying modes   ($k\eta \rightarrow 0^{-}$).
Then,  the corresponding $\Phi$ and $\Psi$ are given by
\bq
\lb{5.15}
\Phi_{k}
= \Psi_{k}  = c_{0} H^{2}\eta \simeq 0,
\eq
as $\eta \rightarrow 0^{-}$.

Therefore, it is concluded that  for any given $\xi$ the de Sitter spacetime is stable against the gravitational
 scalar perturbations.

\section{Conclusions}

In this brief report, we have studied the stability of the de Sitter spacetime  in the framework of HL theory of quantum gravity
with projectability but without detailed balance condition. In contrast to the Minkowski case \cite{WM,KA}, the gravitational
scalar perturbations are stable in the IR limit for any given coupling constant $\xi$. The model is free of ghost
for $\xi \le 0$. Thus, restricting $\xi$ to this range, we can see that both of the ghost and instability problems disappear here
in de Sitter background.

 It should be noted that the analysis given in Section III is valid only in the IR limit, as we dropped all terms proportional
 to $\alpha_{1}$ and $\alpha_{2}$. In particular, in the UV regime, these terms become dominant, and the gravitational
 scalar perturbations will be quite different from the ones given by Eqs. (\ref{5.11}), (\ref{5.13}) and (\ref{5.15}). However,
 by properly choosing the coupling constants $g_{2},\; g_{3},\; g_{7}$ and $g_{8}$,  it can be shown that the solutions are
 free of ghosts and stable for $\xi \le 0$ in the UV regime, too.

Finally, we note that the initial motivation of Horava  was to construct a UV complete theory of quantum gravity, while
in the IR limit it reduces to GR \cite{Horava1,Horava2,Horava3}.  So, in this sense  the HL theory can be considered as an 
ultra-violet-modified gravity. Recently, infrared-modified gravities have also attracted a lot of attention \cite{RT1}. Simple 
examples are the
massive gravity \cite{FP} and the DGP models \cite{DGP}. Although the HL theory and these  infrared-modified gravities
represent modifications of GR in two opposite regimes, surprisingly they face similar problems: ghosts, tachyons, and 
strong couplings. For example,
in massive gravity, the massless spin-0 graviton does not decouple in the massless limit in flat space, the well-known 
vDVZ discontinuity \cite{vDVZ1,vDVZ2}. Although such a discontinuity can disappear   when $M/H \rightarrow 0$
in the de Sitter background, ghosts appear for $ 0 < M^{2} < 2\Lambda/3$  \cite{Hig1,Hig2,KPP}. Ghosts and strong couplings
also happen in DGP models \cite{LPR,DM08.1,DM08.2}. Interestingly, the strong coupling in the DGP models actually 
helps to screen   the spin-0 mode so that the models are consistent with solar system tests \cite{RT2,RT3}.  In addition,
the vDVZ discontinuity also disappears in the anti de Sitter background \cite{KPP,Por}.

~\\{\bf Acknowledgements:} We would like to thank Roy Maartens and Antonios Papazoglou
for valuable discussions and comments.  Part of the work  was supported   by NNSFC   under Grant   10703005 and
No. 10775119 (AZ $\&$ QW).


\end{document}